**Shear Thickening in Polymer Stabilized Colloidal Suspensions**

Joachim Kaldasch* and Bernhard Senge

Technische Universität Berlin

Fakultät III: Lebensmittelrheologie

Königin-Luise-Strasse 22

14195 Berlin

Germany

* author to who correspondence should be sent

**Abstract**

This paper adopts a previously developed activation model of shear thickening, published by the authors, to sterically stabilized colloidal suspensions. When particles arranged along the compression axis of a sheared suspension, they may overcome the repulsive interaction and form hydroclusters associated with shear thickening. Taking advantage of the total interaction potential of polymeric brush coating and van der Waals attraction, the applicability of the activation model is shown within the validity range of a continuum theory. For the comparison with an extensive experimental investigation, where some parameters are not available, the onset of shear thickening can be predicted with realistic assumptions of the model parameters.

# 1. Introduction

Polymeric stabilization, known as "steric stabilization", can be achieved by adsorbing or grafting a polymer onto the surface of colloidal particles (Crowl and Malati (1966), Sato and Ruch (1980), Napper (1983)). Contrary to electrostatic stabilization of colloidal suspensions, steric stabilization is not affected by a variation of the pH value or salt concentration, which is important for many applications. The rheology of sterically stabilized suspensions has been studied intensively (Frith et al (1990), Grover and Bike (1995) Nommensen et al. (1998)). Usually, colloidal suspensions show a highly non-Newtonian behavior even though the medium is a Newtonian liquid. The deviation from Newtonian flow behavior becomes pronounced for high particle volume fractions. At these volume fractions an increase in the applied shear rate may lead to an increase of the apparent viscosity. This phenomenon is referred to as shear thickening.

The theoretical attempts to understand shear thickening in stabilized colloidal suspensions developed into four main directions. Based on light scattering investigations (Hoffman (1974)) shear thickening was related to an order-disorder transition, where an ordered, layered structure becomes unstable above a critical shear rate (Boersma et al. (1990), Hoffman (1998)). The immense progress in computer technology at the end of the last century allowed simulations of a large number of hard spheres under shear. It could be shown that short-range lubrication forces are responsible for the formation of shear induced hydroclusters causing shear thickening. This approach was extended by taking the repulsive interaction of the electric double layer and adsorbed polymer layer into account (Brady and Bossis (1985), Bender and Wagner (1989), Maranzano et al. (2001), Melrose (2003), Melrose and Ball (2004), Krishnamurthy et al. (2005)).

The investigation of rigid spheres in contact lead to the idea that shear thickening can also be understood as a stress-induced transition into a jammed state. Based on a mode-coupling model, where the memory term takes into account the density, shear stress and shear rate, continuous and discontinuous shear thickening were obtained (Cates et al. (1998), Holmes et al. (2003), (2005)).

Recently a fourth approach was proposed by Kaldasch et al. (2008) that is based on an activation model. It suggests that when particle pairs may overcome the mutual repulsion in a suspension perturbed by simple shear, shear thickening occurs due to particle clusters generated along the compression axis of the flowing suspension. This model was originally designed for the prediction of the critical stress of shear thickening in electrically stabilized suspensions. The combined action of the Van der Waals attraction and electrostatic repulsion leads to a interaction potential that exhibits a potential barrier (Verwey and Overbeek (1948), Derjaguin and Landau (1941)). Based on DLVO-and non-DLVO forces the theory shows its applicability in continuous and discontinuous shear thickening (Kaldasch et al. 2009a, Kaldasch et al. 2009b).

The aim of this paper is to generalize the activation approach of shear thickening to sterically stabilized colloidal suspensions. For this purpose we take advantage of the interaction potential of polymeric brush coating colloidal particles and adopt the activation model. On approaching the particle surface the interaction potential of two sterically stabilized particles is considered to be given by a shallow attractive minimum, caused by van der Waals attraction, followed by a steep repulsion due to the polymer brushes. However, if particle cores may be forced in a sheared suspension to approach each other closely, the dominance of van der Waals attraction cannot be neglected. As displayed in Fig. 1 (solid line) the total interaction potential of polymeric brush coating in a continuous theory as derived by Vincent et al. (1986), exhibits a potential barrier into a primary minimum, which allows to apply the activation model of shear thickening to sterically stabilized colloidal suspensions.

In order to show the applicability of the activation model, the theory is used to determine the critical stresses for the onset of shear thickening of an extensive experimental investigation on sterically stabilized PMMA particles in Decalin performed by D' Haene (1995) .

## 2. The Model

We want to consider a concentrated, colloidal suspension of monodisperse spherical particles. In the presence of a stabilizing steric layer the interaction potential, $U(h)$, as a function of the surface-to-surface distance, $h$, can be written as the sum of three contributions:

$$U(h) = U_{vdW}(h) + U_{OS}(h) + U_E(h)$$

(1)

The non-retarded Van der Waals attraction between two spheres has the form:

$$U_{vdW}(h) = -\frac{Aa}{12h}$$

(2)

where $A$ is the effective Hamaker constant determined by the dielectric constants of the solvent-particle combination and $a$ is the particle radius (diameter $d=2a$).

The overlap of polymer brushes for two approaching colloids will result in a local increase of the osmotic pressure due to the increase of the polymer concentration in the overlap region and hence in a repulsive potential. This osmotic potential has the form (Vincent et al. (1986)):

$$U_{OS}(h) = 0 \qquad\qquad\qquad\qquad\qquad\qquad 2L \leq H$$

$$U_{OS}(h) = \frac{4\pi a}{v_L}\phi_p{}^2 k_B T\left(\frac{1}{2}-\chi\right)\left(L-\frac{H}{2}\right)^2 \qquad\qquad L \leq H < 2L$$

$$U_{OS}(h) = \frac{4\pi a}{v_L}\phi_p{}^2 L^2 k_B T\left(\frac{1}{2}-\chi\right)\left(\frac{H}{2L}-\frac{1}{4}-\ln\left(\frac{H}{L}\right)\right) \qquad H < L$$

(3)

where $\chi$ is the Flory-Huggins solvency parameter, $\Phi_p$ is the volume fraction of the polymer within the brush layer of thickness $L$, $v_L$ is the molecular volume of a solvent particle , $k_B$ the Boltzmann constant and $T$ the temperature.

When two particles are closer than a distance equal to $L$, at least some of the polymer molecules are forced to undergo an elastic compression. The elastic deformation of the polymer molecules gives rise to the repulsive potential:

$$U_E(h) = \left(\frac{2\pi a}{Mw}\phi_p L^2 \rho_p k_B T\right)\left(\frac{h}{L}\ln\left(\frac{h}{L}\left(\frac{3-h/L}{2}\right)^2\right) - 6\ln\left(\frac{3-h/L}{2}\right) + 3\left(1-\frac{h}{L}\right)\right)$$

(4)

where $\rho_p$ and $M_w$ are the density and the molecular weight of the adsorbed polymer.

**The critical stress**

Based on an activation model, Kaldasch et. al (2008) suggested that shear thickening occurs in a suspension disturbed by simple shear, if a pair of colloidal particles arranged along the compression axis of the flow may overcome the mutual repulsion of the energy barrier, $U_B$, formed by the two-particle interaction potential:

$$U_B = U(h_{max}) - U(h_0)$$

(5)

where $h_0$ is related to the equilibrium particle surface-to-surface distance, which is governed by the volume fraction $\Phi$ of the solid particles by:

$$h_0(\Phi) \cong 2a\left(\left[\frac{\Phi_m}{\Phi}\right]^{1/3} - 1\right)$$

(6)

while $\Phi_m$ is a maximum packing fraction. The maximum of the interaction potential, at $h_{max}$, can be obtained from

$$\frac{\partial U(h)}{\partial h} = 0; \frac{\partial^2 U(h)}{\partial h^2} < 0$$

(7)

The critical stress of continuous shear thickening in the activation model is determined by

$$\sigma_C = \frac{U_B}{V*}$$

(8)

where $V*$ is the activation volume, which is of the order of the free volume per particle:

$$V* = \frac{4}{3}\pi a^3 \frac{\Phi_m}{\Phi}$$

(9)

We obtain for the critical stress :

$$\sigma_C = \frac{U(h_{max}) - U(h_0)}{\frac{4}{3}\pi a^3 \dfrac{\Phi_m}{\Phi}}$$

(10)

## 3. Comparison with Experimental Investigations

We want to compare our theoretical model with an extensive experimental investigation performed D'Haene (1992) also reported in Frith et al. (1996). Monodisperse PMMA particles of three different particle diameters covered with a $L=9$ nm poly-(12-hydroxy steric acid) (PHSA) layer $(M_w{\sim}500\text{-}1500\ g/mol)$ were dispersed in Decalin.

We want to take advantage from the total interaction potential given by Eq.(1) taking also into account the van der Waals attraction. The Hamaker constant of this particle solvent combination was given $A \cong 0.62 k_B T$ (D'Haene (1992)). The interaction potential can be evaluated if the Flory-Huggins solvency parameter, $\chi$, and the volume fraction of the polymer layer, $\Phi_p$, of the PHSA-layer are known. Since these data are unavailable, we estimate the solvency parameter form an investigation by Fritz et al. 2002, $\chi{\sim}0.485$, and treat the volume fraction of the polymer layer, $\Phi_p$, as an adjustable parameter.

The equilibrium phase diagram suggests that hard sphere colloidal crystals are stable for particle volume fractions $\Phi \geq 0.5$, but there are randomly oriented forming a large scale glasslike structure in equilibrium. Since the volume fractions of the studied suspensions increase the maximum random packing fraction, $\Phi_m=0.64$, some global crystalline order can be expected. For the calculation of the critical stress the maximum packing fraction is therefore assume to be given by a hcp- ordered structure, with $\Phi_m=0.74$. With this assumption the impact of $U(h_0)$ on the potential barrier in Eq.(10) can be completely neglected in the evaluation of the critical stress.

Displayed in Figure 2 are the experimental critical stresses for the onset of shear thickening of the three particle diameters studied by D'Haene (1992) (*d=340 nm, d=690 nm, d=823 nm*). Also shown are the critical shear stresses of shear thickening as evaluated form Eq.(10), using two polymer volume fractions, *$\Phi_p$=0.04* (upper lines) and *$\Phi_p$=0.03* (lower lines). Most of the critical stresses of the smaller particles are located within these two lines. Note, that larger particles are known to exhibit discontinuous shear thickening in concentrated suspensions. However, discontinuous critical stresses are shifted to lower values than predicted by Eq.(10) and become a function of the unknown shear thinning history of the samples (Kaldasch et al. (2009b)). Since the parameter range for polymeric brush coating, is usually between *$0.04<\Phi_p<0.2$,* this result suggests that the activation model is principally applicable to determine the onset of shear thickening in sterically stabilized suspensions, if sufficient information is available.

Krishnamurthy and Wagner (2005) could show that their effective hard sphere model is also able to predict the experimentally obtained onset of shear thickening with realistic assumptions of the unknown parameters. The key idea of this model is the assumption that lubrication forces are responsible for the formation of hydroclusters in a sheared suspension. Taking also into account the hydrodynamic permeability of the polymer coat, they could fit the critical stresses, when they assume that particles approach each other up to a minimum surface-to-surface distance, *$\delta$=3.6 nm.*

Both models fits the experimental data, but the advantage of the present model is the direct relation of the critical stresses to the total interaction potential. From a fit of the dynamic elastic moduli, *$G'_\infty$*, Frith et al. (1996) suggested an effective hard sphere potential of the form:

$$U(h) = Ah^{-B}$$

(11)

where $U(h)$ has the units of joules and the surface-to-surface distance, $h$, has units of nanometers. With $A=1.82\ 10^{-17}$ and $B=3.4$ for the 690 nm particles the interaction potential is displayed in Figure 1 (solid line). The interaction potential as suggested form the activation model displayed by the solid line in Figure 1 coincides very good with the experimentally obtained potential, for $h>2\ nm$, which also indicates the validity of the activation model.

We want to mention another rheological investigation, carried out by Lee et al. (1999) on sterically stabilized silica particles in THFFA, in which they studied the onset of shear thickening as a function of the temperature. They suggested that the onset of shear thickening must be associated with an activation process, since the critical shear rate and the corresponding viscosity are straight lines in an Arrhenius plot. Note, that the presented activation model suggests that the critical stress is substantially independent of the temperature. Calculating the critical stresses of their samples indicate, however, that $\sigma_C$ is indeed temperature-independent as predicted by our model. For example $\sigma_C \approx 30$ Pa, (particle B).

## 4. Conclusions

Previous investigations on shear thickening in sterically stabilized systems treat the colloidal particles as effective hard spheres. The key assumption is that lubrication forces damp the particle approach such that hydrodynamic and steric repulsion prohibits a close touch of the particles. Taking into account the hydrodynamic permeability of the polymer coat Krishnarmuthi and Wagner (2005) showed the applicability of the model with reasonable assumptions of the unknown parameters.

The present work suggests an alternative approach to describe shear thickening in sterically stabilized suspensions. The authors adopt the activation model of shear thickening previously developed for electrically stabilized particles to sterically stabilized colloidal suspensions. The critical shear stresses in this model are directly determined by the steric interaction potential of the colloidal suspensions. A comparison with an experimental investigation on sterically stabilized PMMA particles suspended in Decalin by D'Haene (1992) allows an estimation of the polymer volume fraction of the steric polymer coat. The polymer volume fraction, $\Phi_p$, evaluated from the present activation model is in the expected parameter range for polymeric brush coating. The experimentally suggested effective hard sphere potential could be verified for surface-to surface distances larger than 2 nm. These results indicate the principle applicability of the activation model to shear thickening in sterically stabilized systems.

Note that the maximum of the potential, $h_{max}$, of the sterically stabilized suspensions is located about an Angstrom apart from the surface independent of the particle size ($h_{max}(\Phi_p=0.03)=0.11nm$). Because the model is based on a meanfield interaction potential, the surface roughness of the particles is neglected. The argument of the authors is that the surface roughness may prohibit a permanent capture of the particles into the primary minimum of the van der Waals attraction and causes reversibility of shear thickening.

While the present investigation is limited by estimations of the parameters involved, we want to emphasize that the activation model of shear thickening can be verified quantitatively, if a precise characterization of the interaction potential of the colloidal particles is performed. Therefore experimental investigations with accurately characterized model systems are necessary.

**Figures**

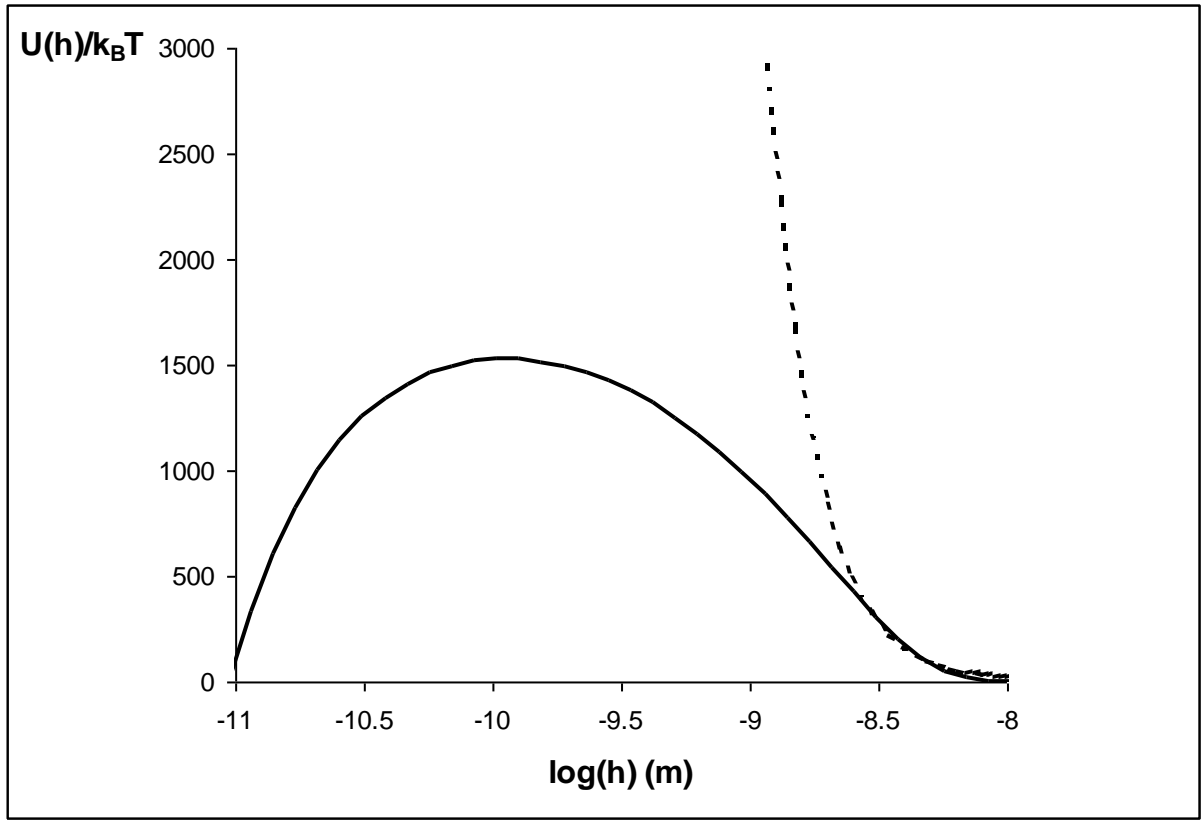

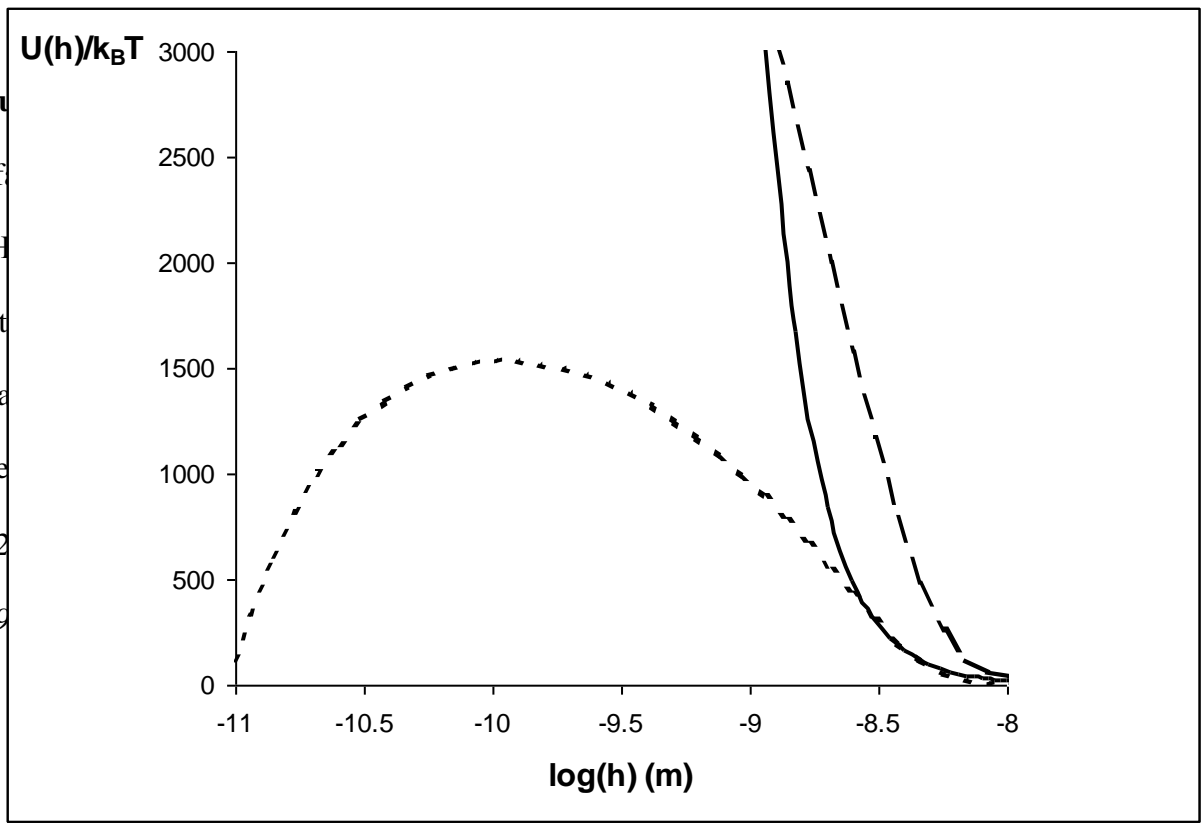

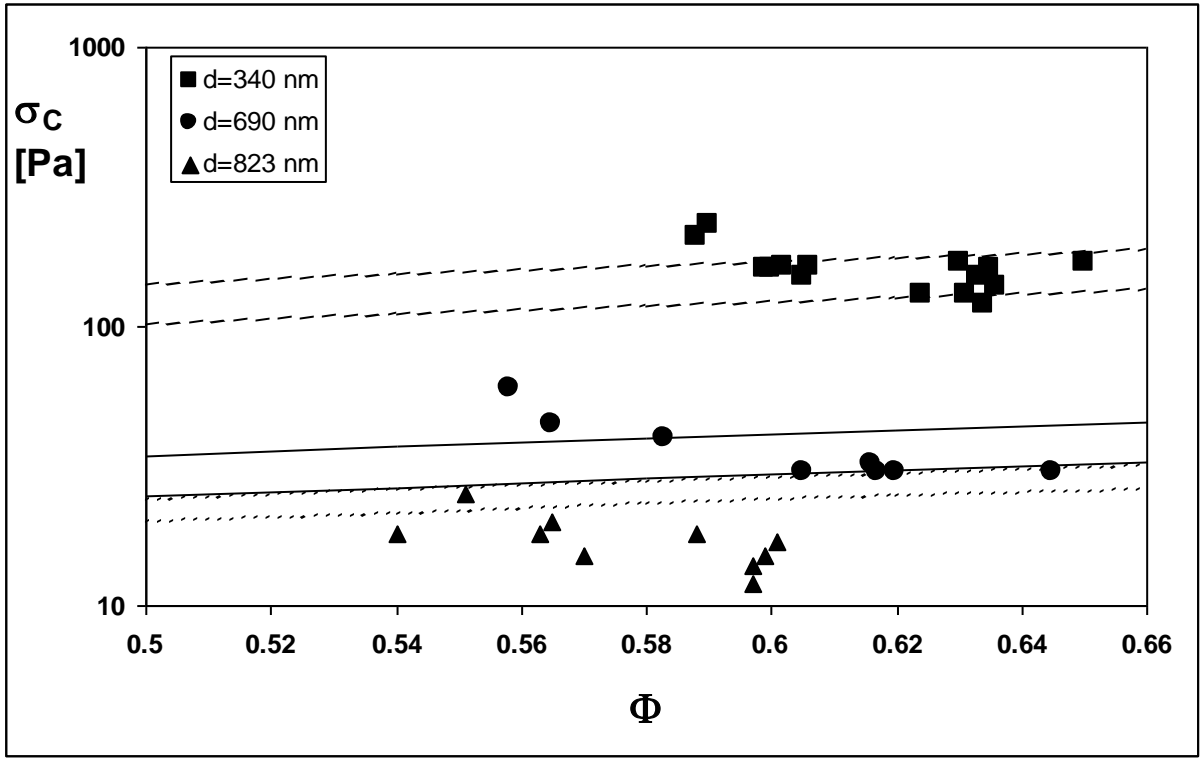

**Figure 2.** Critical stresses for shear thickening as a function of the volume fraction for different PMMA particles in decalin. The dashed lines represent the critical stress for *d=340 nm*, the solid lines for *d=690 nm* and the dotted lines for *d=823 nm*, while the upper lines correspond to $\Phi_p=0.04$ and the lower to $\Phi_p=0.03$. ( D'Haene (1992)).